\documentclass[twocolumn,showpacs,preprintnumbers,amssymb]{revtex4}

\usepackage{graphics}
\usepackage{epsfig} 
\usepackage{hyperref}
\usepackage{graphics}
\usepackage{color}      
\usepackage{graphicx}
\usepackage{graphicx}
\usepackage{dcolumn}

\begin{document}


\title{ Rectified motion of short polymer chain that walks along a ratchet potential that coupled with spatially varying temperature }

\author{  Mesfin Asfaw Taye }
 \affiliation {Department of Physics and Astronomy, California State University\\ Dominguez Hills, California, USA }

\date{\today}

\begin{abstract} 

We explore the transport features of a single flexible polymer chain that walks on a periodic ratchet potential coupled with spatially varying temperature. At 
steady state the polymer exhibits a fast unidirectional motion where the intensity of its current rectification depends strongly on its elastic strength and size.    Analytic and numerical analysis reveal that the steady state transport of the polymer can be controlled by attenuating the strength of the elastic constant.  Furthermore, the stall force at which the chain current vanishes is independent of the chain length and coupling strength.    Far from the stall force the mobility of the chain is strongly dependent on its size and flexibility.   These findings show how the mobility of a polymer can be controlled by tuning system parameters, and may have novel applications for polymer transport and sorting of multicomponent systems based on their dominant parameters .  
\end{abstract}
\pacs{Valid PACS appear here}
\maketitle
 

\section { Introduction} 
There has been much interest in the study of noise-induced transport features of 
biological systems such as polymers and membranes,  with the aim to get  a deeper understanding of how their internal degree of freedoms  affect their dynamics \cite{c1,c2,c3,c4,c5,c6}.  
Often these biological systems contain a large number of different components which are organized in a complex fashion. As a result,  they  exhibit  transport feature that has a nontrivial dependence on their size, flexibility, and the background temperature. 
Previous studies on the dynamics of a flexible polymer chain on a bistable potential showed that the escape rate of the chain is sensitive to the size of the molecule and the strength of interaction between monomers \cite{c7,c8,c9,c10,c11,cc11,c12,c13,c14,c15,c16,c17}. 
Also, the transport features of polymers exposed to a time varying potential reveals the subtle interaction between noise and periodic forces leads to the phenomenon of stochastic resonance (SR) \cite{c18,c19,c20}.   In particular, recent work on the SR of a linearly coupled polymer surmounting  a potential barrier showed that at the resonance temperature the chain undergoes fast unidirectional motion.   This study suggested a novel approach to control the transport properties of important biological molecules such as DNA
\cite{c7,c8,c9,c10,c11,cc11}.

A net unidirectional transport can also be achieved when the polymer is arranged to move along a flashing or rocking ratchet.   Recent studies have also shown that transport in these systems can be controlled by attenuating the chain's flexibility and size \cite{c21,c22,c23}.  This work is consistent with experimental results showing that a Brownian ratchet can lead to fast transport of both particles \cite{c24,c25,c26,c27,c28} and polymers \cite{c29}.   Several groups have also studied the transport properties a monomer in a double-well potential with a spatially varying temperature \cite{m1,m2,m3,m4,m5,m6,m7,m8}.  However, to date there has been no systematic investigation on the transport features of polymer chain in such a system.     Thus in this paper, we consider a flexible polymer moving in a ratchet potential with an external load where the viscous medium is alternatively in contact with the hot and cold heat reservoirs along the space coordinate. The numerical and analytical analyses show that the polymer exhibits a fast unidirectional current where the strength of the current rectification relies not only on the thermal background and load, but also on the coupling strength and size.

In this work, we study the dependence of the velocity of the chain on the coupling strength $k$. For finite $k$, the mobility of chain exhibits a peak and as $k$ further gets increased, the velocity decreases. The velocity of the chain also strictly relies on magnitude of the external load. The velocity decreases  as load increases. It stalls at stall force. As the load further increases, the polymer changes its direction and  its reversed  velocity increases with load. 
Furthermore, our analysis   uncovers that the stall force 
at which the chain current vanishes, is independent of the chain length $N$ and coupling strength $k$. Moreover, we show that the velocity exhibits an optimum value at particular barrier height $U_0$ and as the intensity of background temperature increases, the polymer exhibits a fast unidirectional motion. All of the numerical simulation results are 
justified with exact analytical results in the limit $k \to 0$ and $k \to \infty$.

The paper is organized as follows: In section II, we present the model. In section III,  the role of coupling strength on the mobility of the polymer  is discussed. In section IV, the dependence of the velocity of globular chain  on model parameters is discussed. Section V deals with summary and conclusion.

\begin{figure}[ht]
\centering
{
    \includegraphics[width=8cm]{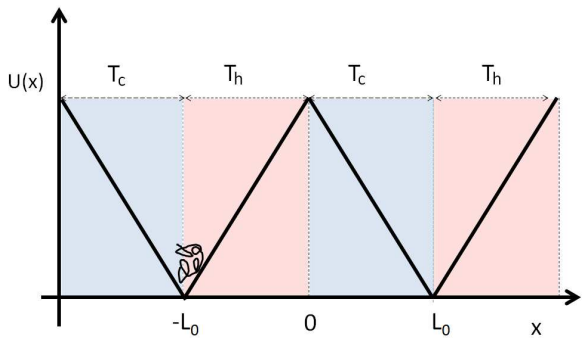}}
\caption{(Color online) Schematic diagram for initially coiled  polymer chain in a piecewise linear bistable potential in the absence of an external load. The potential wells and the barrier top are located at $x=\pm L_{0}$ and  $x=0$, respectively. Due to the thermal background kicks, the polymer ultimately  attains a steady state velocity as long as there is a distinct temperature difference between the hot and cold reservoirs.} 
\end{figure}

\section {The model} We consider a flexible polymer chain of size $N$ which undergoes a
Brownian motion in a one dimensional piecewise linear bistable potential with an external load $U(x)=U_{s}(x)+fx$ where the ratchet potential $U_{s}(x)$ is described by 
 \begin{equation} U_{s}(x)=\cases{
U_{0}\left({x\over L_{0}}+1\right),&if $-L_{0}\le x \le 0$;\cr
   U_{0}\left({-x\over L_{0}}+1\right),&if $0 \le x \le L_{0}.$\cr
   }\end{equation} 
Here,  $U_{0}$ and $2L_{0}$ denote  the barrier height and the width of the ratchet potential, respectively, and where $f$ is the load. The potential has 
 a potential maxima $U_{0}$ at  $x=0$ and potential minima at $x=\pm L_{0}$. In this work, 
 the chain contour length  is taken  to be much less than the characteristic dimension of the ratchet  potential $2L_{0}$.
 The ratchet potential is also coupled with a spatially varying temperature 
 \begin{equation} T(x)=\cases{
T_{h},&if $-L_{0}\le x \le 0$;\cr
  T_{c},&if $0 \le x \le L_{0}$\cr
   }\end{equation} 
 as shown in Fig. 1.  $U_{s}(x)$ and $T(x)$ are assumed to have the same period such that  $U_{s}(x+2L_{0})=U_{s}(x)$ and $T(x+2L_{0})=T(x)$. 

Considering only nearest-neighbor interaction between
the polymer segments (the bead spring model), the Langevin equation that
governs the dynamics of the $N$ beads $(n=1,2,3 ,. . ., N)$ in a highly viscous
medium under the influence of external potential $U(x)$ is given by
 \begin{eqnarray}
\gamma{dx_{n}\over dt}&=&-k(2x_{n}-x_{n-1}-x_{n+1})-{\partial U(x_{n})\over \partial x_{n}}+ \\ \nonumber
&&\sqrt{2k_{B}\gamma T(x_n)}\xi_{n}(t)
\end{eqnarray}
where the $k$ is the spring (elastic) constant of the chain while $\gamma$ denotes the friction coefficient.  $\xi_{n}(t)$ is assumed to be Gaussian  white noise and $k_{B}$ denotes the Boltzmann constant. Hereafter,  we assume  $k_{B}$ to be unity.

To simplify model equations we introduce a dimensionless  load ${\bar f}=fL_{0}/T_{c}$, rescaled temperature ${\bar T}(x)=T (x)/T_{c}$, rescaled barrier height ${\bar U_{0}}=U_{0}/T_{c}$  and rescaled length ${\bar x}=x/L_{0}$. We  also introduced a dimensionless coupling strength   ${\bar k}=kL_{0}^{2}/T_{c}$, $\tau=T_{h}/T_{c}$  and     time ${\bar t}=t/ \beta$ where    $\beta=\gamma L_{0}^2/T_{c}$ denotes the relaxation time. From now on, $\beta$ and $\gamma$ are taken to be unity  and  all the quantities are rescaled (dimensionless) so that the bars will be dropped.

\section{ Flexible  polymer chain} Previous studies  have  shown that  a single monomer   (a Brownian particle) attains a directional motion
when it is exposed to a ratchet  potential coupled with a spatially variable temperature or an external load. For such  a system, the  functional dependence  
for the steady state current $ J$ or the velocity $V$    on the system parameters   is well  explored \cite{m1,m2,m3}.  However, it is not known how these results apply to a chain with several monomers.   Here, we will explore the dependence of the unidirectional chain velocity as a function of key system parameters.

Next in order to understand how the  velocity of the chain responds  to the change  to its   conformational flexibility and variability that arise due to its internal degree of freedoms, we simulate the Brownian dynamics given by Eq. (3) and compute the steady state current.  This result is then averaged over $10^4$ independent simulations.

To analyze further how  the polymer or in general any   linearly coupled system responds to the nonhomogeneous thermal noise  while surmounting a  double-well potential with load, the dependence of the velocity as a function of the different system parameters  is explored. The numerical and analytical analyses reveal that  the polymer exhibits a unidirectional current where the strength of the current rectification  relies not only on the thermal background kicks and load but it has also a nontrivial dependence on its coupling strength  and size.  It is found that in the absence load $f=0$, the chain maintains  a positive current as long as     a distinct temperature difference between the hot and cold reservoirs is retained; i.e,  $T_{h} >T_{c}$, $V>0$. For isothermal case, a one dimensional negative current can be achieved providing $f \ne 0$. In general when $T_{h} >T_{c}$ and $f \ne 0$, the polymer exhibits intriguing  transport features. Figures 2 plots 
 the velocity   $V$  as a function of $k$  for fixed external load $f=0.0$ and $f=4.0$, respectively. 
The numerical results  exhibits that in the limit $k \to  \infty$, $V$ goes to the velocity of a  rigidly coupled polymer (dashed blue line) that evaluated via Eq. (8); when $k \to 0$, $V$ approaches  to the velocity of a single Brownian particle (  dashed blue line) that evaluated using Eq. (8).  The same  figure depicts that the chain retains a higher velocity  at $k=0$  than a globular chain ($k \to \infty$). Another crucial feature such model system is that the  chain internal degree of freedoms has the capacity to enhance  the chainÓ³ speed. As a result, the current does manifest a noticeable optimal peak  at a certain optimal $k$.  
 \begin{figure}[ht]
\centering
{
    \includegraphics[width=8cm]{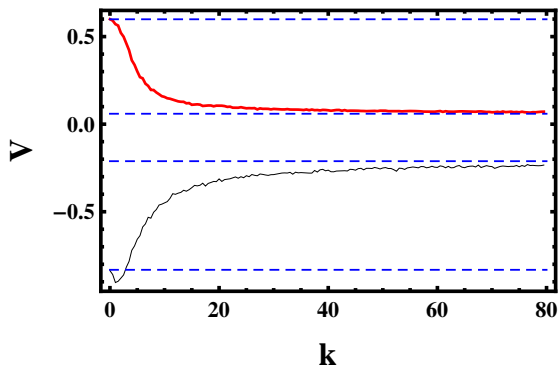}}
\caption{(Color online) The velocity $V$  as a function of  $k$ for parameter choice $f=0.0$ (red solid line) and $f=4.0$ ( black solid line).  In the figure, other parameters are fixed as 
 $N=2.0$,  $U_{0}=6.0$ and $\tau=2.0$. The dashed blue lines are from the exact analytic results (Eq. (8)) both in the limit of $k\to \infty$ and $k\to 0$} 
\end{figure}

 Next via numerical simulations, we explore the dependence of the chain stall force $f'$ on its internal degree of freedoms. 
 Surprisingly the numerical analysis reveals that   for the  flexible polymers   with finite $k$,    the stall force is still independent of $N$ which is in agreement to the exact analytical result for the  globular chain (see Eq. (9)).
  At this point we want to stress that the external load dictates the direction of the particle flow. 
When  $f<f'$, the net current is positive and while on the contrary for $f>f'$, the current flows from the cold  to the hot reservoirs. It worth noting that a larger  polymer  moves   sluggishly than a smaller chain  as long as $f \ne f'$. At stall force $f=f'$, the  polymer  will have zero velocity  regardless   its size.   In Fig. 3a, we plot $V$ as a function of $f$. In the figure, the green solid lines stand the plot for $V$ in the limit of $k\to 0$ (top)  and $k \to \infty$. The dotted lines are analyzed from the simulations for given values of $k=0$, $k=8.0$ and $k=25.0$ (globular chain) from the top to bottom, respectively.
As depicted in Fig. 3a, for polymer with finite $k$, current reversal occurs at $f'=2.0$ for parameter choice  $U_{0}=6.0$ and  $\tau=2.0$    regardless of the magnitude of $k$ revealing that the coupling strength is not a relevant control parameter to alter the direction of polymer's current.  On the other hand 
   Figure 3b depicts the plot of $V$ as a function of $U_{0}$ for a parameter choice $f=0.3$ and $\tau=2.0$.  As shown in the figure, the velocity for the polymer  monotonously  increases  with  $U_0$ and attains a maximum  value at a particular optimum  barrier height $U_{0}^{opt}$. Further increasing in $U_{0}$ leads  to a smaller $V$. At   $U_{0}^{opt}$, the chain retains a maximum speed. The same figure shows that the velocity increases when $k$ decreases. $U_{0}^{opt}$ also strictly relies on $k$; when $k$ decreases, $U_{0}^{opt}$ increases. Furthermore, our analysis  exhibits that  
  the transport property of the  chain  also strictly relies on the temperature difference between the hot and cold baths. When the magnitude of the rescaled temperature  $\tau$ steps up, the tendency for the polymer in the hot bath to reach the top of the ratchet potential increases than the chain in the cold reservoir. This leads to an increase in the current or velocity.
  \begin{figure}[ht]
\centering
{
    \includegraphics[width=6cm]{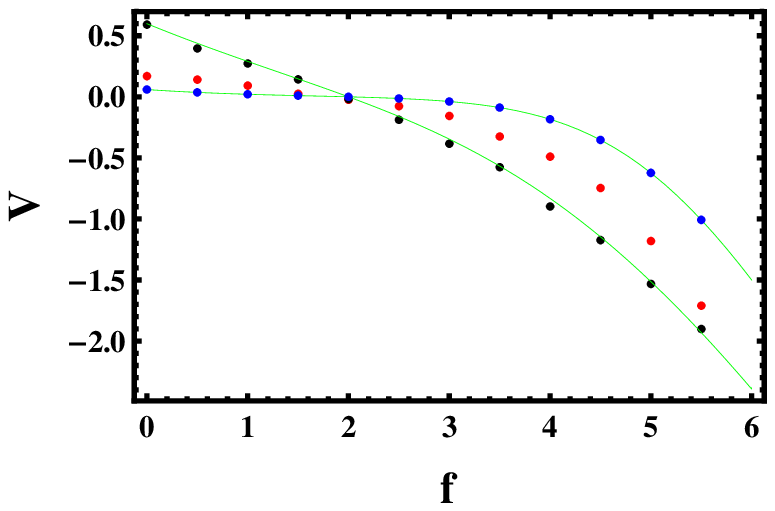}}
\hspace{1cm}
{
    \includegraphics[width=6cm]{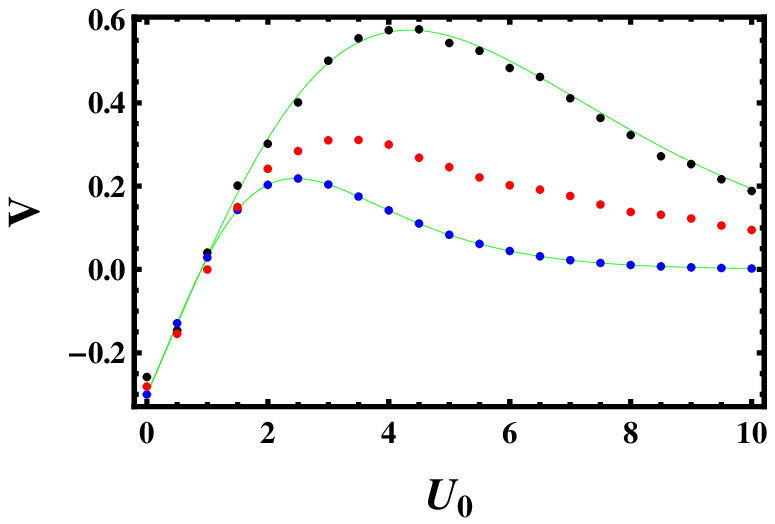}
}
\caption{ (Color online)(a) The velocity  $V$  as a function of $f$ for the  parameter values of $U_{0}=6.0$, and   $\tau=2.0$. The green solid lines stand the plot for $V$ in the limit of $k\to 0$ (top)  and $k \to \infty$. The dotted lines are analyzed from the simulations for given values of $k=0$, $k=8.0$ and $k=25.0$ (globular chain) from the top to bottom, respectively. (b) The velocity  $V$ as a function of $U_{0}$ for parameter choice $k=0$, $k=5.0$ and   $k=25.0$ ( compact chain), from top to bottom. We also fixed $f=0.3$ and  $\tau=2.0$; dotted line stands for the simulation results  while green solid lines are form analytic prediction. } 
\label{fig:sub} 
\end{figure}

\begin{figure}[ht]
\centering
{
    \includegraphics[width=8cm]{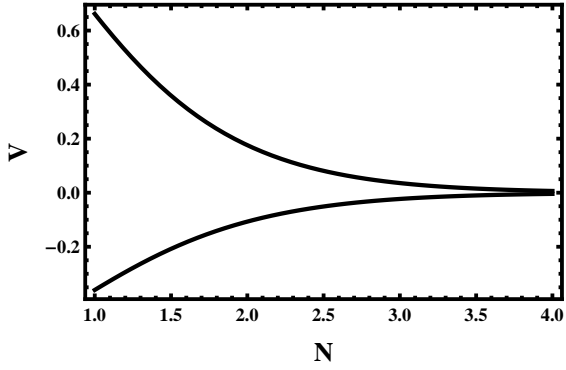}}
\caption{(Color online) The velocity $V$  as a function of  $N$ for parameter choice $f=0$ and $f=2.0$ from top to bottom.  In the figure, other parameters are fixed as $k=20$,  
  $U_{0}=4.0$ and $\tau=2.0$. } 
\end{figure}
The dependence of the velocity on chain's length  is also investigated for parameter choice $f=0$ and $f=2.0$ from top to bottom.  In the figure, other parameters are fixed as $k=20$,  
  $U_{0}=4.0$ and $\tau=2.0$. (see Fig. 4). The figure depicts that when the load is not strong enough, the polymer attains a positive current while for large load, the system exhibits a current reversal. In both cases, the chain velocity monotonously decreases  as the chain length decreases.

\section{Globular polymer chain}

In order gain a deeper insight into this  finding, it is instructive to compute the velocity for globular polymer as well as a single Browinian. 
  In order to rewrite the Langevin equation  for compact polymer or rigid polymer ($k \to \infty$) in terms of the center of mass motion, let us add the $N$ Langevin equations (Eq. (3)) to get
 \begin{eqnarray}
{d \over dt}{(\sum _{i=1}^{N}x_{i})} = -\sum _{i=1}^{N}{\partial U(x_{i})\over  \partial x_{i}}+\sum _{i=1}^{N}\sqrt{2\gamma T(x_i)}\xi_{i}(t).
 \end{eqnarray}
When  a compact  polymer of size $N$ hops on the ratchet potential,  each  monomer experiences  the same  force along the reaction coordinate. Hence 
the effective Langevin equation  for the center of mass motion $x_{cm}=(x_{1}+x_{2}+ ...x_{N})/N$ can be written as 
 \begin{eqnarray}
N{d x_{cm} \over dt} = -N{\partial U(x_{cm})\over  \partial x_{cm}}+\sqrt{2\gamma T(x)}(\xi_{1}(t)+...+\xi_{N}(t)).
 \end{eqnarray}
 From fluctuation-dissipation relation 
 \begin{eqnarray}
\left\langle (\xi_{1}(t)+...+\xi_{N}(t))(\xi_{1}(t)+...+\xi_{N}(t))\right\rangle=N\left\langle \xi(t)^2 \right\rangle
 \end{eqnarray}
 which implies that we can substitute  $(\xi_{1}(t)+...+\xi_{N}(t))$ by $\sqrt{N}\xi(t)$. After some algebra Eq. (5) converges to 
 \begin{equation}
{dx_{cm} \over dt}= -{\partial U(x_{cm})\over  \partial x_{cm}}+\sqrt{2\gamma T_{cm}(x)}\xi(t)/\sqrt{N}.
  \end{equation} 
  The corresponding steady state current $J$  can be exactly evaluated  using the same approach as the work \cite{m3}. After some algebra, we find a closed form expression for the steady state current
  \begin{equation}
J= -{\varsigma_{1}\over \varsigma_{2}\varsigma_{3}+\varsigma_{4}\varsigma_{1}}
  \end{equation} 
  where the expressions  for $\varsigma_{1}$, $\varsigma_{2}$, $\varsigma_{3}$ and $\varsigma_{4}$ are given as 
 $
  \varsigma_{1}=e^{a-b}-1$,
  $\varsigma_{2}={N\over a \tau}\left(1-e^{-a}\right)+{N\over b}e^{-a}\left(e^{b}-1\right)$, 
  $\varsigma_{3}={1\over a}\left(e^{a}-1\right)+{1\over b}e^{a}\left(1-e^{-b}\right)$. 
   The parameter  $\varsigma_{4}$ is given by $\varsigma_{4}=\epsilon_{1}+\epsilon_{2}+\epsilon_{3}$ where 
   $
  \epsilon_{1}={N\over \tau}\left({1\over a}\right)^2\left(a+e^{-a}-1\right)$, 
 $\epsilon_{2}={N\over ab}\left(1-e^{-a}\right)\left(e^{b}-1\right)$, 
  $\epsilon_{3}=N\left({1\over b}\right)^2\left(e^{b}-1-b\right)$. 
   Here $a=N\left(U_{0}+f\right)/\tau$  and  $b=N\left(U_{0}-f\right)$.
   The corresponding velocity is given by $V=2J$.
   In the limit $f \gg U_{0}$ and large $N$ we have $a \approx Nf/\tau$, $b \approx -Nf$, $\varsigma_{1} \approx \exp [Nf/ \tau +Nf]$, $\varsigma_{2} \approx 1/f$, $\varsigma_{3} \approx \exp[ (Nf/ \tau +Nf)]/Nf$,
  $ \epsilon_{1}\approx1/f$,  $\epsilon_{2}\approx 0$ and  $ \epsilon_{3}\approx1/f$. Substituting these values, we get 
$
J\approx -f/2.
 $  
 Furthmore, the exact analytical result for the  globular chain uncovers that the stall force 
 \begin{equation}
f'={U_{0}(\tau-1)\over (\tau+1)}
  \end{equation}
 at which the chain current  vanishes, is independent of the chain length $N$.

In the absence of external load $f=0$, the steady state current (Eq.(8)) converges to
  \begin{equation}
J={NU_{0}^2\over 2(1+\tau)}\left[{1\over e^{NU_{0}\over \tau}-1}-{1\over e^{NU_{0}}-1}\right].
  \end{equation}
  For small $U_{0}$, it is straight forward to show 
   $
J\approx{U_{0}\over 2}\left({\tau-1 \over \tau+1}\right).
 $
  On the other hand for large $U_{0}$ and $\tau$,
  one approximates Eq. (10) as 
   $
J \approx{NU_{0}^2\over 2(1+\tau)}e^{-NU_{0}\over \tau}.
$

Closer look at the Fig. 2 once again reveals that the chain retains a higher velocity  $V(0)$ at $k=0$ than the velocity $V(\infty)$ of a globular chain ($k \to \infty$). Particularly, as the size of the chain increases,   the gap between $V(0)$ and $V(\infty)$ increases.  To analyze the chain size dependence further, we have computed the ratio for the velocity of  a single particle to globular polymer utilizing Eq. (8). As exhibited in Fig. 5, $V$  is a nontrivial function of $N$; the polymer with small $k$ retains considerably higher velocity than a rigid dimer. This signifies that attenuating the strength of the elastic constant results in a polymer that can be transported fast. This can be notably appreciated
by taking   the  velocity ratio between a single and globular polymer in high barrier limit which  is given as 
     \begin{equation}
     {V(0)\over V(\infty)}={e^{(-1+N)(fL_{0}+U_{0})\over \tau}\over N}
  \end{equation}
  where in the limit $f \to 0$,  ${V(0)\over V(\infty)}={\exp^{(-1+N)(U_{0})\over \tau}\over N}$. This can be retrieved using our previous calcualtions since for large $U_{0}$, 
   $
J \approx{NU_{0}^2\over 2(1+\tau)}e^{-NU_{0}\over \tau}.
$
\begin{figure}[ht]
\centering
{
    \includegraphics[width=8cm]{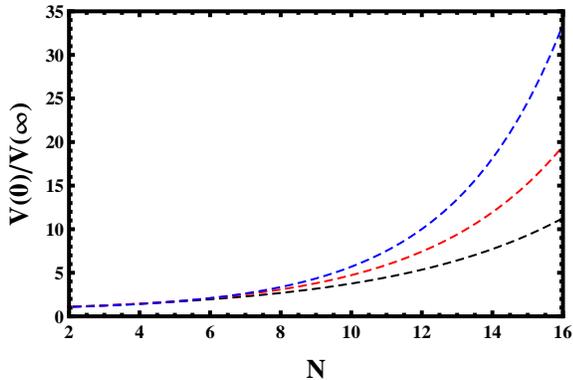}}
\caption{The ratio $V(k\to 0)/V(k\to \infty)$  as a function of  $N$ for parameter choice $f=0.0$ (black solid line), $f=0.5$ (red solid line), $f=1.0$ (blue solid line). Other parameters are fixed as   $U_{0}=2.0$ and $\tau=8.0$.} 
\end{figure}

The central results of this paper also indicates   the occurrence   a direct
relationship between the flexibility of a macromolecule and
its transport properties. Hence, we expect that in general
this relationship can be applied to control the transport of
molecules by modulating their flexibility. 
Modifying the
flexibility of a macromolecule can be achieved in a variety
of ways. Experimentally, 
the flexibility of the chain   can be manipulated  in a variety of ways.  For instance, the flexibility of
proteins can be altered by ligand binding \cite{m10}.
  The
elasticity of the DNA molecule  can be also  strengthen  by introducing external
charges \cite{m11}.  Thermal 
and chemical denaturation also alter  the flexibility
of biological molecules since hydrogen
bond breaking  leads to an increase in the rotational degrees of freedom of atoms and thereby  increases the
macroscopic flexibility of the molecule \cite{m12,m122}.

 \section { Summary and conclusion} 
We  study the transport and  response properties  of a single flexible polymer moving in a ratchet potential with an external load where the viscous medium is alternately in contact with inhomogeneous temperature along the reaction coordinate. As long as the system is far from equilibrium, we show that each monomer of the chain exhibits a fast unidirectional current  
where the strength of the current rectification relies not only on the thermal background kicks and load but it has also a nontrivial dependence on its coupling strength and size. 

The numerical and exact analyses indicate that  the stall force 
is independent of the chain length $N$ and coupling strength $k$. It is also shown that a flexible chain  retains a higher velocity than a less flexible polymer revealing that a chain with a desired speed can be fabricated by  
 attenuating the strength of the elastic constant. The chain's flexibility  can be modified  through ligand binding \cite{m10},
by introducing external
charges \cite{m11} and via 
chemical denaturation 
\cite{m12,m122}.

In conclusion, in this work, we present a pragmatic model system which 
 not only serves as a basic guide on how to transport the polymer fast to specific region but also has novel applications for  binding kinetics, DNA amplifications and sorting of multicomponent systems based on their  dominant parameters

{\it Acknowledgment.\textemdash} 
This work was supported in part by the National Heart,
Lung, and Blood Institute (Grant No. R01HL101196). I would like to thank Yohannes Shiferaw for the interesting discussion I had. I would like also to thank Mulu Zebene for the constant support.

\end{document}